\documentclass[conference,a4paper]{IEEEtran}
\IEEEoverridecommandlockouts
\usepackage{cite}
\usepackage{amsmath,amssymb,amsfonts}
\usepackage{algorithmic}
\usepackage{graphicx}
\usepackage{textcomp}
\usepackage{xcolor}
\usepackage{color,soul}
\usepackage{graphicx, caption, subcaption}
\usepackage{upgreek}
\def\BibTeX{{\rm B\kern-.05em{\sc i\kern-.025em b}\kern-.08em
    T\kern-.1667em\lower.7ex\hbox{E}\kern-.125emX}}

\usepackage{comment}

\begin{document}

\title{28 GHz Wireless Channel Characterization for a Quantum Computer Cryostat at 4 Kelvin
\thanks{Authors gratefully acknowledge funding from the European Commission via projects with GA 101042080 (WINC) and 101099697 (QUADRATURE).}
}

\author{\IEEEauthorblockN{
Ama Bandara\IEEEauthorrefmark{1},   
Viviana Centritto Arrojo\IEEEauthorrefmark{1},   
Heqi Deng\IEEEauthorrefmark{2},    
Masoud Babaie\IEEEauthorrefmark{2},      
Fabio Sebastiano\IEEEauthorrefmark{2},      
Edoardo Charbon\IEEEauthorrefmark{3},  \\    
Evgenii Vinogradov\IEEEauthorrefmark{1},      
Eduard Alarc\'on\IEEEauthorrefmark{1},      
Sergi Abadal\IEEEauthorrefmark{1}      
}                                     
\IEEEauthorblockA{\IEEEauthorrefmark{1}Nanonetworking Center in Catalunya, Universitat Polit\`{e}cnica de Catalunya, Barcelona, Spain}
\IEEEauthorblockA{\IEEEauthorrefmark{2} 
Delft University of Technology, 2628 CD Delft, The Netherlands}
\IEEEauthorblockA{\IEEEauthorrefmark{3} 
Ecole Polytechnique F\'ed\'erale de Lausanne (EPFL), Switzerland}

 \IEEEauthorblockA{ \emph{*ama.peramuna@upc.edu} }
}

\maketitle

\begin{abstract}

The scalability of quantum computing systems is constrained by the wiring complexity and thermal load introduced by dense wiring for control, readout and synchronization at cryogenic temperatures. To address this challenge, we explore the feasibility of wireless communication within a cryostat for a multi-core quantum computer, focusing on wireless channel characterization at cryogenic temperatures. We propose to place on-chip differential dipole antennas within the cryostat, designed to operate at 28 GHz in temperatures as low as 4~K. We model the antennas inside a realistic cryostat and, using full-wave electromagnetic simulations, we analyze impedance matching, spatial field distribution, and energy reverberation due to metallic structures. The wireless channel is characterized through measured channel impulse response (CIR) across multiple receiver antenna positions. The results demonstrate potential for reliable short-range communication with high Signal-to-Noise Ratio (SNR) and limited sensitivity to positional variation, at the cost of non-negligible delay spread, due to significant multipath effects. 
\end{abstract}


\section{Introduction}

The fast-evolving landscape of quantum computing requires practical solutions to move beyond the Noisy Intermediate-Scale Quantum (NISQ) era \cite{preskill2018quantum}. Achieving quantum advantage for classically intractable problems demands systems with millions of qubits \cite{Ladd2010}. However, scalability is constrained, among other aspects, by the vertical interconnects linking cryogenic quantum processors and room-temperature control electronics \cite{das2024chip}. These dense wire bundles introduce spatial congestion and thermal load which degrades qubit coherence.


One approach to alleviate the vertical wiring bottleneck is to integrate control and readout circuits closer to the qubits. In this regard, advances in cryo-CMOS electronics \cite{charbon2021,patra2018} enable local control, readout, and synchronization at 4~K. This approach is complementary to the recent emergence of modular quantum architectures \cite{Brown2016, Isailovic2006, Monroe2014, alarcon2023scalable} which address scalability by interconnecting multiple Quantum Processing Units (QPUs) \emph{horizontally} at cryogenic temperatures. 

\begin{figure}[!t]
\centering
\includegraphics[width=0.95\columnwidth]{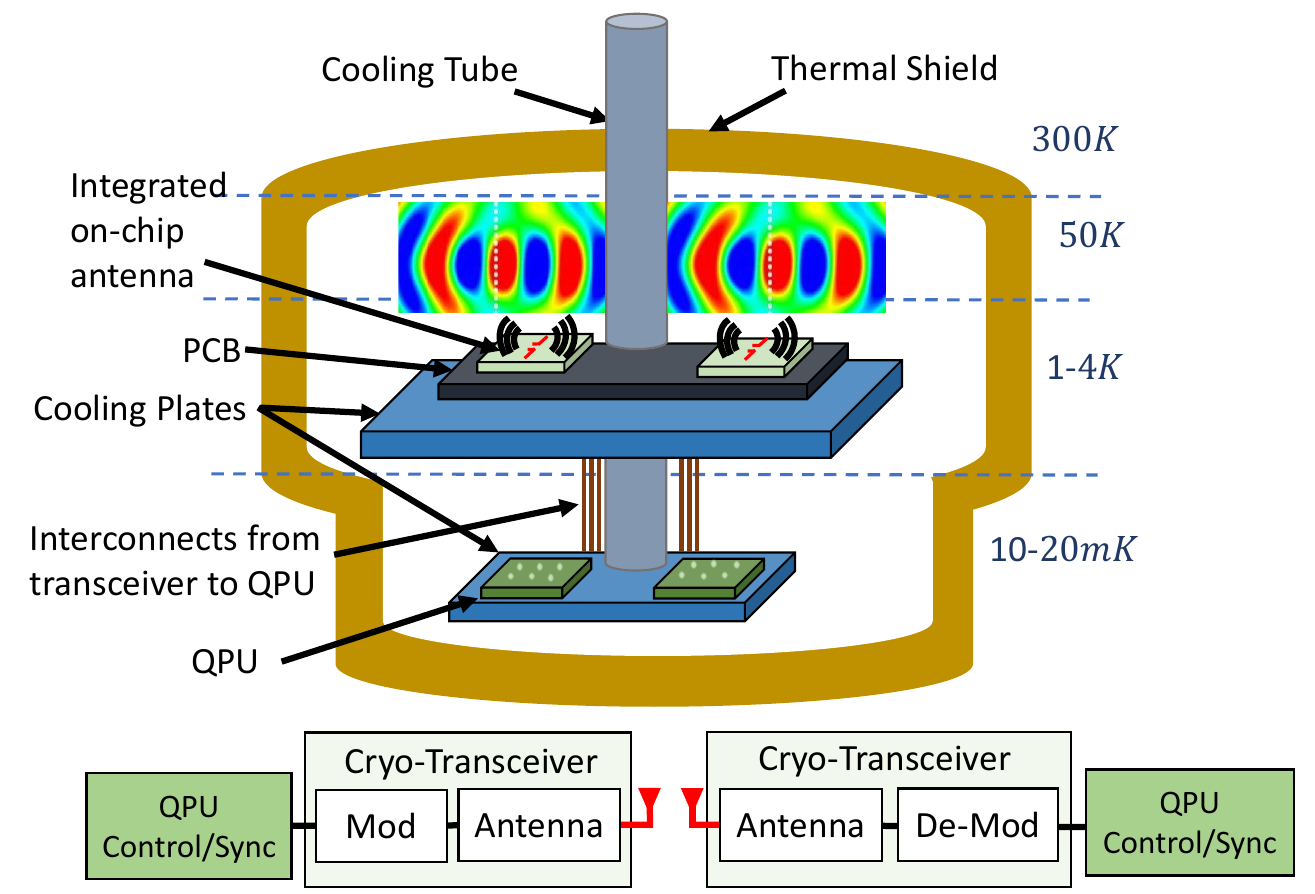}
\vspace{-0.1cm}
\caption{Integrated on-chip differential dipole antennas inside the cryostat, which are positioned above the QPUs. The cryo-transceivers are connected to the QPUs through wires, while the wireless plane is fulfilling the need for classical interconnects (control/synchronization) in a modular architecture.} 
\label{fig:systemmodel}
\vspace{-0.3cm}
\end{figure}

The convergence of modular quantum architectures with the local control circuit approach, albeit promising, poses new interconnect requirements. This is because the control and readout circuits of each QPU shall be interconnected to synchronize operation and orchestrate quantum state transfers across the QPUs, possibly across multiple dies or even boards. This communication can be performed through wired interconnects, yet it could be slow and inefficient if the signaling needs to propagate across multiple boards.

\begin{figure*}[!t]
\centering
\begin{subfigure}[t]{0.33\textwidth} 
\includegraphics[width=1\textwidth]{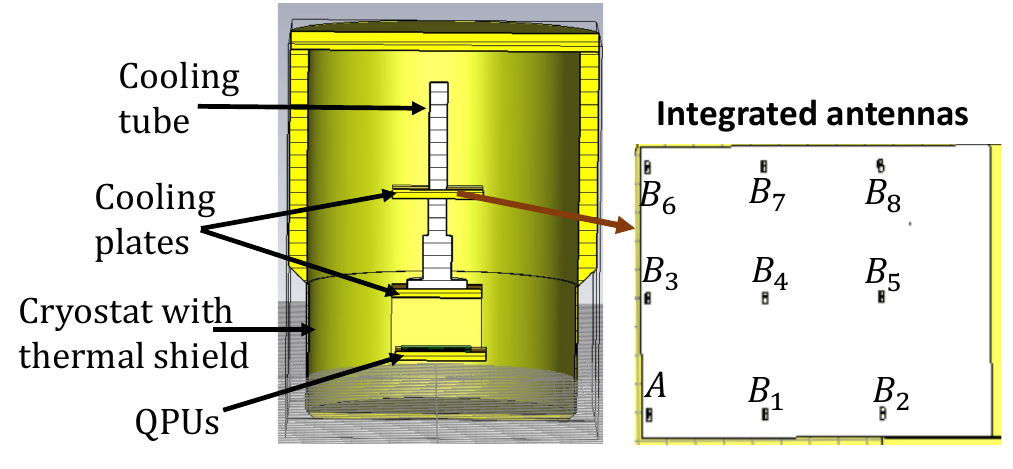}
\caption{\label{fig:antennapositions}}
\end{subfigure}
\begin{subfigure}[t]{0.28\textwidth} 
\includegraphics[width=\textwidth]{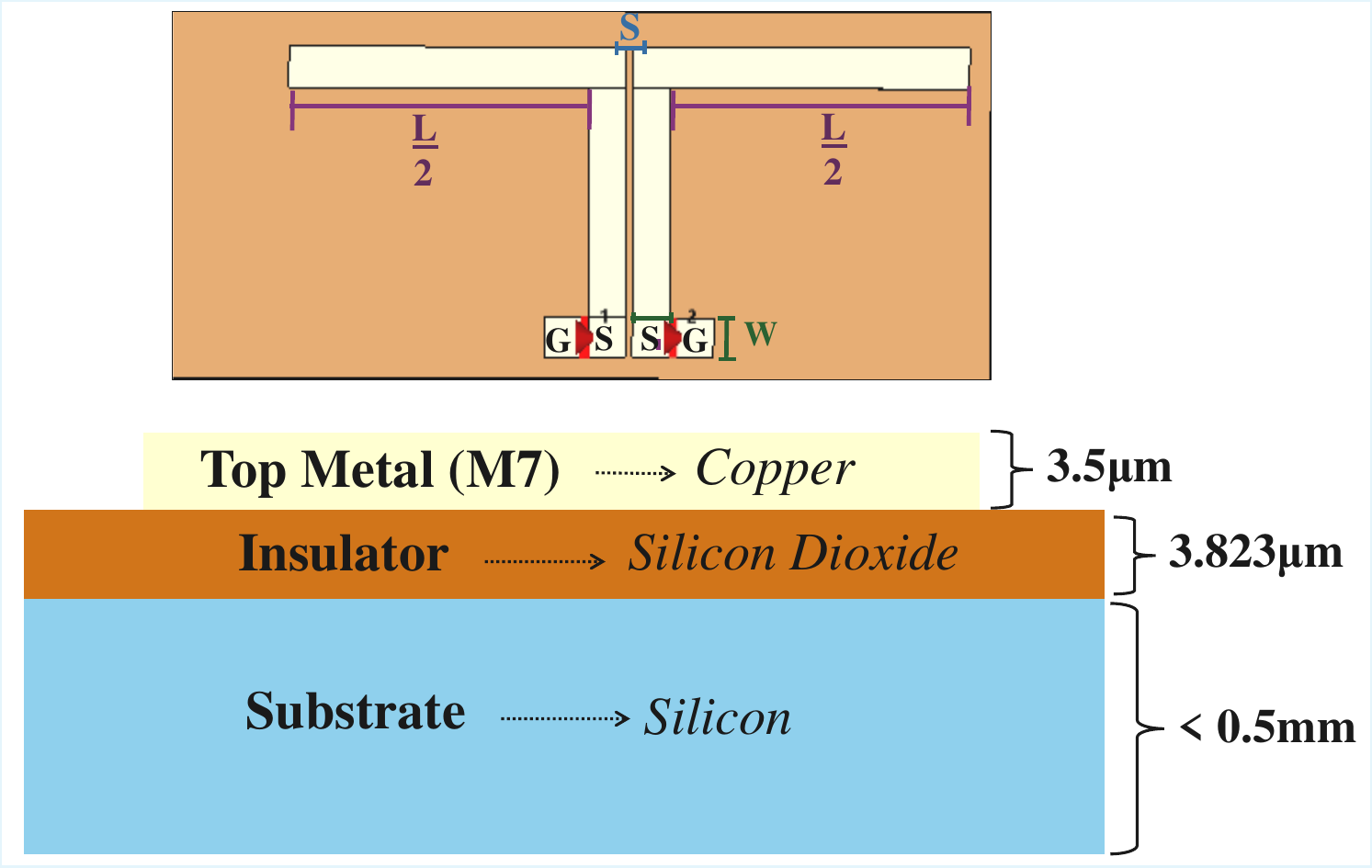}
\caption{\label{fig:dipoledesign}} 
\end{subfigure}
\begin{subfigure}[t]{0.3\textwidth} 
\includegraphics[width=\textwidth]{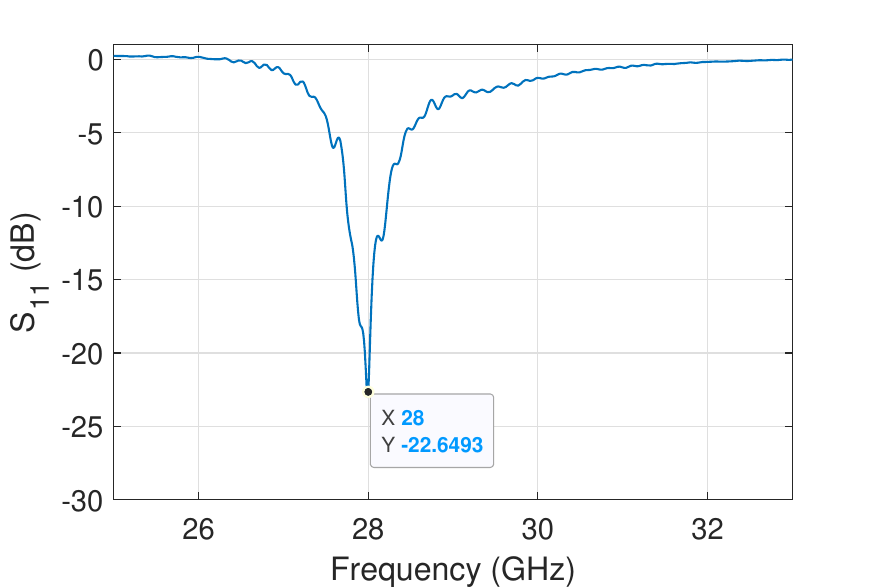}
\caption{\label{fig:s11cryostat}}
\end{subfigure}
\vspace{-0.1cm}
\caption{Simulation setup implemented in CST MWS. (a) Cross-section of the simulation model of the cryostat, together with a detailed map of the evaluated antenna positions within the boundaries of 4~K temperature level. (b) Top-view and cross-section of the differential-fed dipole antenna design, including details on the thickness of the top metal, insulator and substrate layer. (c) Reflection coefficient (dB) of the on-chip differential dipole within the cryostat, showing operation at 28 GHz.}
\label{fig:snrcap}
\vspace{-0.2cm}
\end{figure*}


To further enhance scalability, complementing wired horizontal interconnects with wireless links (Fig.~\ref{fig:systemmodel}) offers global reconfigurability and low latency while operating with minimal thermal impact~\cite{escofet2023interconnect}. Furthermore, due to the minimal thermal noise at cryo-CMOS RF-transceivers, the Signal-to-Noise Ratio (SNR) of such communication is expected to be high, even for very low transmission powers.

Implementing wireless links within a cryostat introduces new electromagnetic challenges. The quantum computer is physically enclosed within a cryostat chamber, which contains power cables, metallic plates for separate cooling stages, metallic cooling tubes, heat shielding materials as the enclosure, and other solid structures with varying material properties. These components create a reverberant environment with multipath propagation that can perturb nearby qubits. Such an enclosed environment resembles the chip-scale wireless communication approach, the channel characterization aspects of which have been surveyed in \cite{abadal2020} from mmWave to optical frequencies. Moreover, prior research has explored various on-chip antenna designs~\cite{Rayess2017, bellanca2017integrated, timoneda2018channel, pano2020, timoneda2018channel, Gutierrez2009, pan2011, may2010}. However, the design and characterization of on-chip antennas for cryo-CMOS environments remain largely unexplored.

In recent literature, cryo-wireless interconnects have been gaining interest. In \cite{kazim2023}, wireless signal transmission using a wideband 8–12 GHz patch antenna inside a dilution refrigerator was investigated, but only assessed at 77~K. In \cite{Wang2025}, a wireless vertical communication system between 4~K and 300~K at 260 GHz has been explored. The system consisted of a horn antenna positioned at the high temperature level, pointing towards the lower part of the cryostat, where a set of on-chip antennas at 4~K would backscatter readout signals. The on-chip antennas showed improved efficiency from 38\% at 300~K to 67\%-97\% due to the considerable improvement of copper's electrical conductivity. 
However, to the best of our knowledge, an analysis for horizontal links inside a quantum computer cryostat at 4~K temperature does not exist.

To bridge this gap, we implement a realistic cryostat model~\cite{bluefors_quantum_2025} in CST Microwave Studio~\cite{CST} with the main goal of characterizing the wireless channel for horizontal communication across QPUs. Following frequency planning guidelines, we design a 28 GHz on-chip differential dipole antenna separated from the spin-qubit operation band at 16 GHz \cite{Philips2022}. The antenna is optimized for operation at 4 K and compatible with cryo-CMOS transceivers. As main contribution, we then characterize the resulting wireless channel inside the cryostat to evaluate its feasibility, considering the effects of enclosure geometry and material properties at cryogenic temperatures. To this end, we measure the channel impulse response (CIR), from which we infer signal strength, delay spread, and SNR at cryogenic conditions.


The rest of the paper is organized as follows. 
 Section~\ref{sec:method} describes the design and placement of the on-chip differential dipole antenna, the cryostat characteristics, and metrics of study. In Section~\ref{sec:results}, we present the antenna and channel characterization results. Finally, Section~\ref{sec:conclusion} summarizes the key contributions and outlines directions for future research. 



\section{Methodology} 
\label{sec:method}
In this section, we provide a detailed description of the (i) design of the differential dipole antenna in cryogenic temperature, (ii) integration of the proposed antenna design into the cryostat, (iii) the channel characterization and performance metrics used for evaluation.

\subsection{Cryo-CMOS Antenna Model}
The proposed antenna is a differential dipole integrated in cryo-CMOS technology \cite{charbon2021, patra2018}, designed to operate at 28~GHz within the cryostat. The choice of differential dipole antenna is based on its noise immunity and ability to suppress common-mode interference through differential feeding \cite{shamim2008}. The choice of the frequency is backed up by the relatively small wavelength, ease of on-chip integration, the maturity of designs for 5G applications, and sufficient distance to resonance frequency of spin-qubits \cite{Philips2022}. Full details of the design are given in \cite{centritto2025}.

The dipole, illustrated in Fig.~\ref{fig:dipoledesign}, comprises two co-linear copper arms deposited on a Silicon Dioxide (SiO$_2$) layer. Initially, the resonant length $L$ was estimated as $L=\tfrac{\lambda_{0}}{2\sqrt{\varepsilon_{eff}}}$,
where $\lambda_{0}$ denotes the free-space wavelength and $\varepsilon_{eff} = \tfrac{\varepsilon_{r}+1}{2}$ represents the effective permittivity of the antenna, with $\varepsilon_{r}$ being the permittivity of the substrate \cite{PlanarDipole2013}, \cite{Abbosh2012}. The final optimal dimensions were refined through iterative adjustments of the length of the dipole's arms.

The antenna is fed through a differential feeding network consisting of two coplanar microstrip lines (MLSs) located on the SiO$_2$ layer. The width ($W$) and spacing ($S$) of the lines were calculated based on the substrate properties and dimensions. Finally, the MLSs terminate in a ground-signal-signal-ground (GSSG) pad configuration.

The proposed on-chip antenna is embedded in a CMOS layered stack, as shown in Fig.~\ref{fig:dipoledesign}. The copper dipole is positioned on the top metal layer, M7, beneath which lies a thick SiO$_2$ layer serving as an insulating medium, followed by a bottom Si substrate \cite{cheema2013}. The antenna design incorporates the Cu conductivity, Si conductivity, Si permittivity, and SiO$_{2}$ permittivity as $2.9\times10^{8}$~S/m, $4.26\times10^{-7}$~S/m, 11.45 and 3.9 respectively, at cryogenic temperature \cite{patra2020, krupka2006}. These values differ from the properties at room temperature and affect the antenna performance as detailed in \cite{centritto2025}.

\subsection{Cryostat Model}
The cryostat, outlined in Fig.~\ref{fig:antennapositions}, is a cylindrical enclosure which is used to isolate the quantum computer from the external noise and interference, while maintaining the adequate thermal profile. The outer layer of the cryostat is consisting of shielding material for magnetic interference and inner layer is with a thermal shield. Based on the refrigeration system used to maintain the temperature, there are different temperature levels at each stage of the cryostat that reach 10-20 mK at the bottom, where the QPUs are housed. In addition to the wired interconnects, each temperature stage contains cooling plates, cooling tubes, and other metallic components, which contribute to multipath signal reflections and signal attenuation due to their material properties and geometric obstructions.
 
The proposed differential dipole antennas are placed at the 4~K temperature together with the cryo-CMOS control circuits \cite{centritto2025}, located above the QPUs. The in-plane separation distance between the different antennas varies from $3.7\lambda$ to $10.8\lambda$ as shown in Fig.~\ref{fig:antennapositions}. Vertically, there are three cooling plates connected to the central cooling tube, where two plates resemble a stage to house the QPUs and one plate is place above the mK temperature level. The thermal shield exhibits reflective properties for the electromagnetic waves which is made of metallic alloys \cite{bluefors_quantum_2025}. This placement helps to reduce communication latency while minimizing thermal interference caused by RF radiation. 
The size of the cryostat is 30~cm diameter with 70~cm height, with the top-middle and middle-bottom plate separation being 15~cm and 10~cm, respectively, based on the range of standard cryostat dimensions used for quantum computing applications, e.g., in \cite{bluefors_quantum_2025}. The antenna plane is on top of the top plate.

\subsection{Performance Metrics}
Next, we analyze the performance metrics which we used for the channel characterization. We consider a single transmitter (TX) antenna $A$ and multiple receiver (RX) antennas $B_1$-$B_8$, one level above the QPUs as shown in Fig.~\ref{fig:antennapositions}. Using the full-wave time-domain solver of CST, we excite the TX with an impulse and record the resulting CIR, denoted as $h(t)$, at the RX antennas. The wireless links are evaluated for each pair of TX-RX antennas that result in different separation distances and alignments along the two directions in the chip plane, relative to the fixed position of the TX antenna $A$.

\subsubsection{Signal to Noise Ratio}\label{SNR}
To quantify the amount of energy received at RX with extensive multipath propagation in the enclosed cryostat environment, we assume a unitary transmitted energy. Then, we measure the SNR of the wireless link as
\begin{equation}
SNR=\frac{P_{TX}\int |h(\tau)|^{2} d\tau}{N},
\end{equation}
this is, dividing the total received signal strength of the message sent with a transmit power of $P_{TX}$, integrated over the entire time period, by the noise power $N$.


The power of the thermal noise is characterized by, $K_BTB$ where $K_B$ is the Boltzmann's constant, $T$ is the temperature in Kelvin and $B$ is the finite bandwidth of the system. However, this formula is an approximation for low frequencies and room temperatures of the more general Nyquist equation \cite{nyquist1928}
\begin{equation}
N=\frac{hf_{c}B}{\exp({{\frac{hf_{c}}{K_{B}T}}})-1},
\end{equation}
where $f_{c}$ is the operating frequency of the antenna and $h$ is Planck's constant. 

\subsubsection{Delay Spread}\label{delayspread}
Using the dispersive wireless channel response $h(\tau)$, we evaluate the root mean square (RMS) delay spread as a measure of the multipath effects, as given by:
\begin{equation}
 DS_{RMS}=\sqrt{ \frac{ \int |h(\tau)|^{2} (\tau - \bar{\tau})^2 d\tau }{ \int |h(\tau)|^{2} } }, 
\end{equation}
where the mean delay is $\bar{\tau} = \frac{ \int |h(\tau)|^{2} \tau d\tau }{ \int |h(\tau)|^{2}  }$.

\section{Performance Evaluation }
\label{sec:results}




To characterize the wireless channel using differential dipole antennas in a cryostat environment, we implement the cryostat model and antennas in CST Microwave Studio \cite{CST} using the directives of the previous section.  

\subsection{On-Chip Antennas Inside the Cryostat}

The integrated on-chip antenna inside the cryostat is simulated using two discrete ports placed on GSSG pads, excited by impulse signals. The antenna design is fine-tuned to achieve efficient radiation and 
50 $\Omega$ reference impedance to the cryostat environment \cite{centritto2025}. The optimized dipole length has been determined to be $L_{\text{opt}} = 3.06$~mm. As shown in Fig.~\ref{fig:s11cryostat}, the reflection coefficient of the transmitter antenna $S_{11}$ exhibits a resonance close to 28 GHz with a minimum value of -22~dB, indicating a proper impedance matching and enhanced radiation performance within the cryostat.

\begin{figure}[!b]
\centering
\includegraphics[width=\columnwidth]{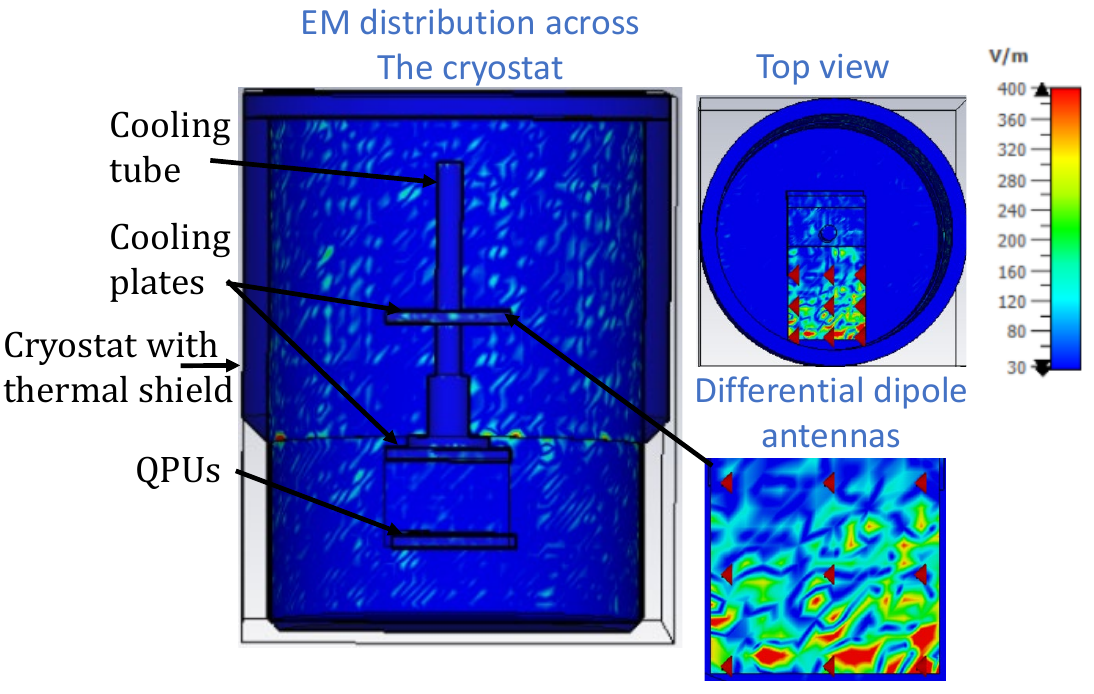}
\caption{Spatial distribution of the electrical field across the cryostat as observed in the cross-section, general top view, and top view at the plane of the antennas.} 
\label{fig:spatialcryostat}
\end{figure}

\begin{figure*}[!t]
\centering
\begin{subfigure}[t]{0.31\textwidth} 
\includegraphics[width=1\textwidth]{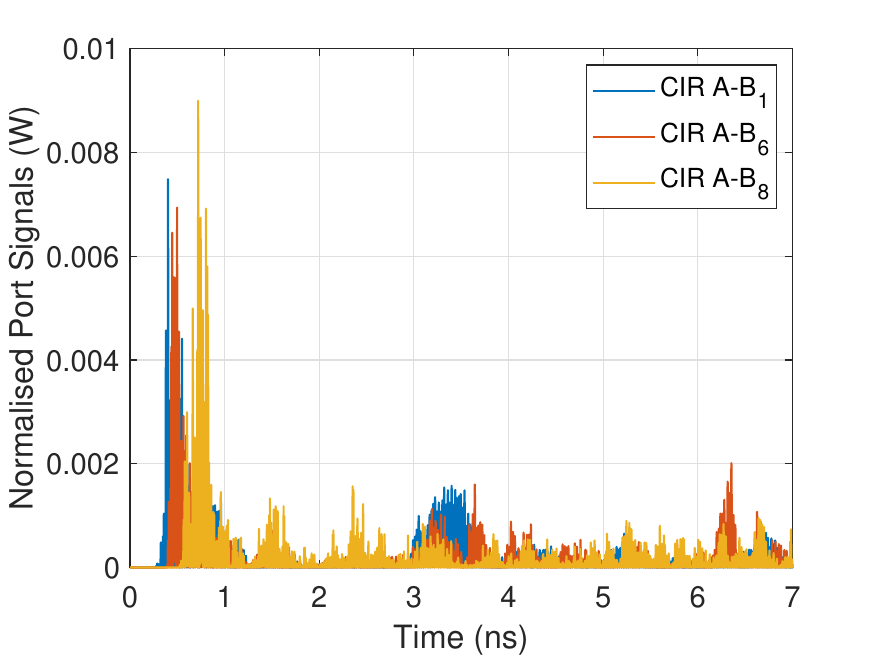}
\caption{\label{fig:cir}}
\end{subfigure}
\begin{subfigure}[t]{0.31\textwidth} 
\includegraphics[width=1\textwidth]{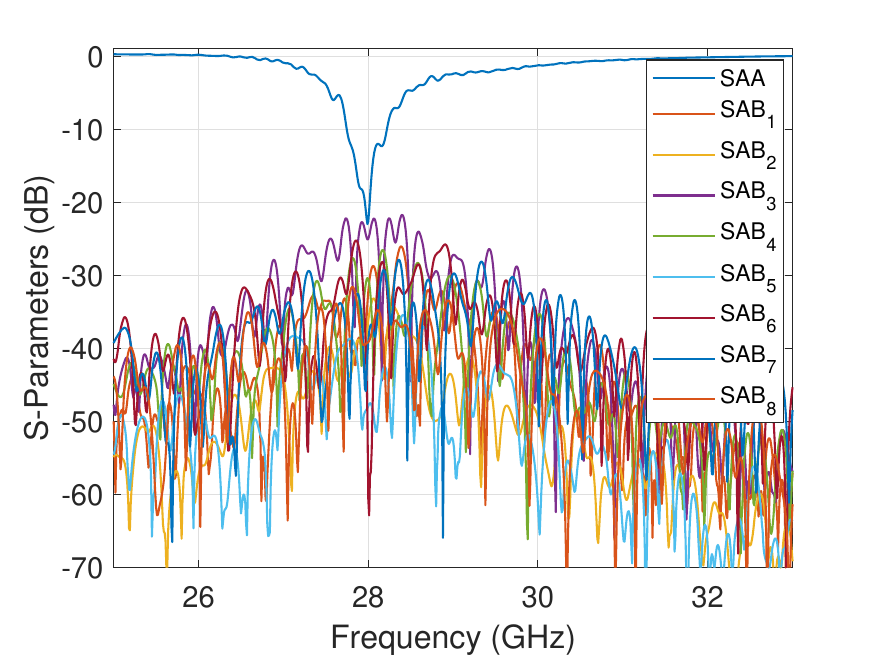}
\caption{\label{fig:s12cryostat}}
\end{subfigure}
\begin{subfigure}[t]{0.31\textwidth} 
\includegraphics[width=1\textwidth]{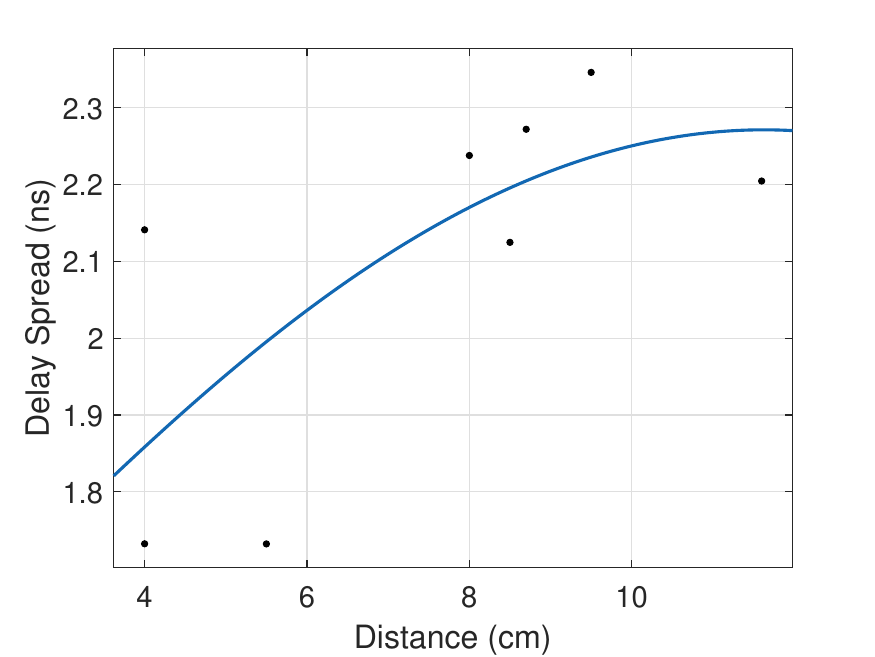}
\caption{\label{fig:ds}}
\end{subfigure}
\begin{subfigure}[t]{0.31\textwidth} 
\includegraphics[width=1\textwidth]{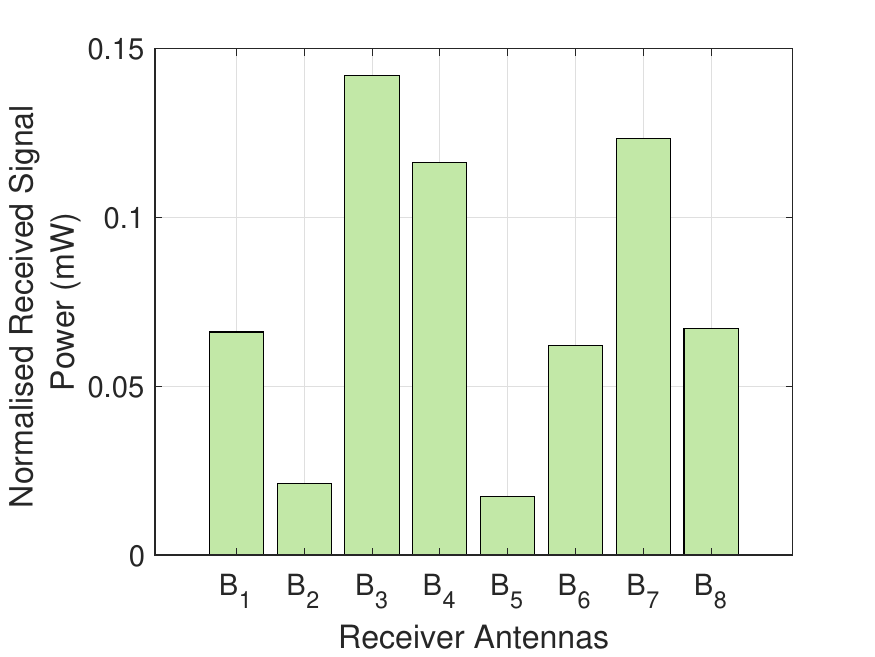}
\caption{\label{fig:receivedpowercryostat}}
\end{subfigure}
\begin{subfigure}[t]{0.31\textwidth} 
\includegraphics[width=1\textwidth]{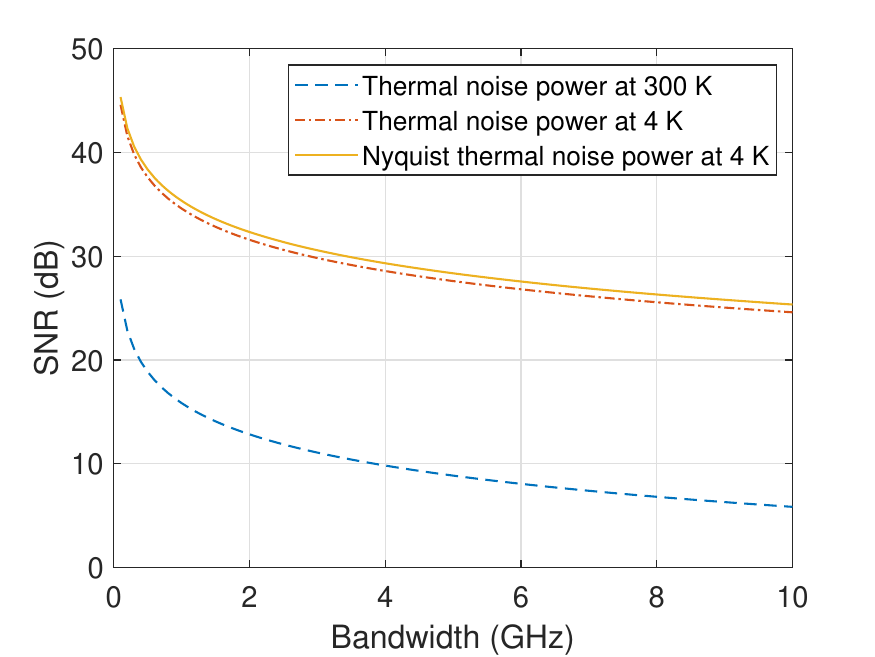}
\caption{\label{fig:snrcryostat}}
\end{subfigure}
\vspace{-0.1cm}
\caption{Characterization of the wireless channel in a 4~K cryostat environment at 28 GHz. (a) Measured power delay profile  of the wireless links $A-B_{1}, A-B_{6},A-B_{8}$ with $P_{TX}=0$~dB (b) Return loss of transmitter $A$ and $S_{21}$ of the different antenna for $B_{1-8}. $ (c) Delay spread over distance for the wireless links $A-B_{1-8}$ with differential feeds. (d) Received power at intended antenna positions of $B_{1-8}$ from a single transmission. (e) SNR as a function of bandwidth assuming $P_{TX}= -30$~dBm.}
\label{fig:snrcap}
\vspace{-0.2cm}
\end{figure*}

The spatial field distribution of the antenna is observed in Fig.~\ref{fig:spatialcryostat}. The antennas are placed on top of the second cooling plate to minimize the RF effects to the QPUs. The figure shows how the energy is distributed primarily through the PCB that holds the antennas, within its same temperature level. Due to reflections and scattering effects, the signal energy also travels to other temperature levels, with some radiation arriving to the cooling plate below. Significant amount of radiation arrives to the antennas that are co-lateral, either because the differential feed radiates or because of common substrate effects or reflections from the tubes or enclosure present in the environment.

\subsection{Wireless Channel}
Multipath effects are observed in the measured CIR as shown in Fig.~\ref{fig:cir}, corresponding to the links $A-B_{1}, A-B_{6}, A-B_{8}$. These effects are due to the enclosed nature of the cryostat and the radiation impairments introduced by the silicon substrate of the on-chip antenna, which result in different delay profiles. Fig.~\ref{fig:s12cryostat} shows the reflection and transmission coefficients (S-parameters) of the channels. Most links show a similar value of the transmission coefficient at 28 GHz, which underscores the influence of reflections and other phenomena reducing the impact of distance on the link response.

In Fig. \ref{fig:ds}, the RMS delay spread of the CIR is obtained as detailed in prior sections. It is observed that the delay spread is relatively high, between 1.7 ns and 2.2 ns, leading to coherence bandwidths $B \sim DS^{-1}$ in MHz. The delay spread initially increases with distance, indicating that the main \emph{direct ray} for communication weakens while distant reflections gain importance. At higher distances, the delay spread is slightly reduced as all reflections seem to be equally important. This is a usual behavior in enclosed environments \cite{abadal2020}.


The received power at each antenna position from $B_{1}$ to $B_{8}$ is shown in Fig.~\ref{fig:receivedpowercryostat}, illustrating how much energy is radiated towards each receiver antenna. Such localized power concentrations could potentially affect neighboring QPUs if the radiated power exceeds the allowable thermal dissipation threshold within the cryostat. As the separation between antennas ranges between $3.7\lambda$ and $10.8\lambda$ depending on the receiver antenna position, the received power varies despite of the distance.
This is mainly due to energy reverberation within the cryostat. The tradeoff, then, lies in the energy-delay correlation since, owing to this reverberant behavior, both energy variation and delay spread can be expected.

In Fig.~\ref{fig:snrcryostat}, the SNR is measured as detailed in Section \ref{SNR} and represented for different thermal noise power models in varied bandwidths, at a transmit power of –30 dBm. 
The thermal noise added at the receiver 
is inherently low at high frequencies and cryogenic temperatures due to low carrier mobility at low temperatures. Yet, when the bandwidth increases, the noise is proportionally increasing, thereby reducing the SNR. Furthermore, it is observed that there is a significant increase of SNR in 4~K compared with room temperature due to low noise figures. 


\section{Conclusion}
\label{sec:conclusion}

In this paper, we have analyzed the feasibility of wireless communication inside a cryostat through integration of on-chip differential dipole antenna operating at 28~GHz down to 4~K. Through full wave simulations, the proposed antenna demonstrated efficient radiation with good impedance matching ($S_{11} \approx -22$~dB) in a cryogenic environment. Channel characterizations have shown that even with energy reverberation and multipath effects, the wireless links maintain a high SNR with low noise. Moreover, the received power at each antenna location is similar, revealing that the spatial energy is dispersed and an adequate received power is obtained despite the distance for wireless receivers. In the near future, we plan to explore the behavior of transmissions under different antenna design strategies and analyze the classical–quantum interconnects while assessing their error rates and interference characteristics, so as to enhance scalability and reliability in multi-core quantum systems. Further, we also aim to assess the amount of radiation that may impact on the QPUs and hence interferece with the qubit operation, to then develop protection measures.

\bibliographystyle{IEEEtran}
\bibliography{IEEEabrv,bib}

\end{document}